\documentclass[aps,prl,twocolumn,nofootinbib,
]{revtex4-2}

\usepackage[english]{babel}

\usepackage{hyperref}
\usepackage{url}
\usepackage{setspace}
\usepackage{amsmath,amssymb,amscd,braket}
\usepackage{amsfonts}
\usepackage{color}
\usepackage{graphicx}
\usepackage{xcolor}

\newcommand{\R}{\mathbb{R}}
\newcommand{\wt}[1]{\widetilde{#1}}

\def\p{\partial}
\newcommand{\bea}{\begin{eqnarray}}
\newcommand{\eea}{\end{eqnarray}}
\newcommand{\be}{\begin{equation}}
\newcommand{\ee}{\end{equation}}
\newcommand{\ba}{\begin{align}}
\newcommand{\ea}{\end{align}}

\begin{document}
\title{An embedding formalism for CFTs in general states on curved backgrounds}
\author{Enrico Parisini}
\email{e.parisini@soton.ac.uk}
\affiliation{Mathematical Sciences and STAG Research Centre, University of Southampton, Highfield, Southampton SO17 1BJ, UK}
\author{Kostas Skenderis}
\email{k.skenderis@soton.ac.uk}
\affiliation{Mathematical Sciences and STAG Research Centre, University of Southampton, Highfield, Southampton SO17 1BJ, UK}
\author{Benjamin Withers}
\email{b.s.withers@soton.ac.uk}
\affiliation{Mathematical Sciences and STAG Research Centre, University of Southampton, Highfield, Southampton SO17 1BJ, UK}

\begin{abstract}
We present a generalisation of the embedding space formalism to conformal field theories (CFTs) on non-trivial states and curved backgrounds, based on the ambient metric of Fefferman and Graham. The ambient metric is a Lorentzian Ricci-flat metric in $d+2$ dimensions and replaces the Minkowski metric of the embedding space. It is canonically associated with a $d$-dimensional conformal manifold, which is the physical spacetime where the CFT${}_d$ lives. We propose a construction of CFT${}_d$ $n$-point functions in non-trivial states and on curved backgrounds using appropriate geometric invariants of the ambient space as building blocks. This captures the contributions of non-vanishing 1-point functions of multi-stress-energy tensors, at least in holographic CFTs. We apply the formalism to 2-point functions of thermal CFT, finding exact agreement with a holographic computation and expectations based on thermal operator product expansions (OPEs), and to CFTs on squashed spheres where no prior results are known and existing methods are difficult to apply, demonstrating the utility of the method.  
\end{abstract}

\maketitle
It is important to understand quantum field theory in curved backgrounds and nontrivial states. This is the case both for purely theoretical reasons and also because it has many applications in a wide range of physical scenarios, from condensed matter systems at finite temperature, to out of equilibrium physics ({\it e.g.} the quark-gluon plasma), to semiclassical black hole physics and cosmological observables. 

A special class of quantum field theories are CFTs. They appear as fixed points under Renormalisation Group flow, and at second-order phase transitions, meaning that they are ubiquitous in nature. They are also of considerable theoretical interest since they enter the anti-de Sitter/conformal field theory (AdS/CFT) correspondence.

The kinematical constraints of CFTs in (conformally) flat spacetimes in vacuum have been solved long ago \cite{Polyakov:1970xd, DiFrancesco:1997nk}. In particular both 2-point and 3-point functions of primary operators are fixed by conformal symmetry up to constants, while higher-point functions are fixed up to functions of cross-ratios. The analogue of these results for CFTs in curved spacetime and nontrivial states is not available and the purpose of this \emph{Letter} is to fill in this gap. In particular we will propose a framework for solving the kinematical constraints due to Weyl invariance and apply it to scalar 2-point functions.

\paragraph{The embedding space.}
The embedding space formalism takes advantage of the fact that the conformal group $SO(1,d+1)$ in $d$ dimensions coincides with the Lorentz group in $(d+2)$ dimensions \cite{Dirac:1936fq, Boulware:1970ty, Costa2011,Costa2011a}. Imposing conformal invariance on CFT observables on any $d$-dimensional conformally flat background ${\cal M}$ simply reduces to  demanding Lorentz invariance on the embedding space $\R^{1,d+1}$.  This construction is realized by mapping $\cal M$ to a projective section of the lightcone $X^M X_M=0$ in Minkowski space\footnote{Throughout this letter, small Latin indices $i,j \dots$ are $d$-dimensional, small Greek indices $\mu,\nu\dots$  denote the $d+1$ directions on hyperbolic spaces while capital Latin letters $M,N, A, B \dots$ denote the $d+2$ embedding or ambient directions. We will sometimes write $X^M$ as the following triplet $X^M = (X^0, X^i, X^{d+1})$.}: taking the CFT background to be flat space $g_{(0)ij}=\delta_{ij}$, each point $x^i \in {\cal M} = \R^d$ can be mapped to a null ray in $\R^{1,d+1}$ according to
\begin{equation}\label{EspaceLCEmbedding}
    X^M(t,x^i) = t \left( \frac{1+x^j x^j}{2}, x^i, \frac{1-x^j x^j}{2}\right),
\end{equation}
with $t \in \R$ and where $t=1$ corresponds to the isometric embedding of $\R^d$.
A linear $SO(1, d+1)$ transformation maps null rays to null rays and this is equivalent to standard conformal transformations on $\R^{d}$.
Considering scalar correlators, the only building blocks that can be constructed on $\R^{1,d+1}$ out of the positions of the insertions are the scalars $X_{ij} = -2 \, X_i \cdot X_j$ of dimension $-2$. As a consequence, using a scaling argument 2-point functions of scalar operators $O$ of dimension $\Delta$ are fixed to
\begin{equation}
\label{EspaceTwopoint}
    \braket{O(X_1) O(X_2)} = \frac{C_\Delta}{(X_{12})^\Delta},
\end{equation}
where $C_\Delta$ is a theory specific constant.
After projecting back to $\R^d$, 
$\left. X_{12} \right|_{t=1}= |x_1 - x_2|^2$, 
recovering the known expression.
With similar reasoning one can efficiently obtain the form of tensorial correlators and higher-point functions.

We now generalise the embedding space formalism to apply to the case of CFTs in non-conformally flat backgrounds $g_{(0)}$ and non-trivial states by using the geometrical construction known as the \emph{ambient space} \cite{Fefferman85, Fefferman:2007rka}. The ambient space allows one to impose the kinematical constraints of Weyl invariance \emph{in lieu} of full conformal symmetries by finding Weyl invariants on a $d$-dimensional manifold as diffeomorphism invariants in $d+2$ dimensions.\footnote{Other uses of the ambient space in physics include (but are not limited to) higher spin theories and holographic anomalies \cite{Grigoriev:2011gp,Joung:2013doa,Bekaert:2017bpy,Grigoriev:2018mkp,Curry:2014yoa,Gover:2011rz,RodGover:2012ib}.}

\paragraph{The ambient space.}
To introduce the ambient space construction we first rewrite the flat metric on $\R^{1,d+1}$ with the new coordinates $(t,\rho,x^i)$,
\begin{equation}\label{AspMApToEsp}
    Y^M(t,\rho, x^i) = t \left( \frac{1-2\rho+x^2}{2}, x^i, \frac{1+2\rho-x^2}{2}\right).
\end{equation}
The result is the Minkowski$_{d+2}$ metric in \emph{ambient form}
\begin{equation} \label{EspaceAform}
    \eta_{MN} dY^M dY^N = 
    2\rho dt^2 + 2t dt d\rho + t^2 \delta_{ij} dx^i dx^j.
\end{equation}
The surface $\rho=0$ where $X^M(t,x^i)=Y^M(t,0,x^i)$ describes the lightcone and in this limit we recover \eqref{EspaceLCEmbedding}. The region $\rho<0$ corresponds to the future and past of the lightcone, while $\rho>0$ covers the points with a space-like separation from the origin. The coordinate $t= Y^+ = Y^0+Y^{d+1}$ defines various sections of the lightcone.
\eqref{EspaceAform} admits a homothety $T = t\partial_t$, a null vector which geometrises scaling transformations of the $d$-dimensional theory.

The ambient space generalises \eqref{EspaceAform} so that it applies to a general background $g_{(0)}$ and general states. There are two key ingredients. First the generalised spacetime should possess a homothety $T=t\partial_t$ and a nullcone structure at $\rho=0$. Second it should be Ricci flat. The most general metric satisfying these conditions up to diffeomorphisms is of the form \cite{Fefferman85,Fefferman:2007rka}
\begin{equation} \label{AMetricTRho}
    \tilde{g} = 
    2\rho dt^2 + 2t dt d\rho + t^2 g_{ij}(x,\rho) dx^i dx^j,
\end{equation}
for which the Ricci tensor $\wt{R}_{MN}=0$. When the Riemann tensor vanishes this reduces to the embedding space \eqref{EspaceAform}. 
Given a boundary metric $g_{ij}(x,0) = g_{(0)ij}(x)$ the Ricci-flat condition can be solved in the neighbourhood of $\rho=0$,
\begin{equation}
    g(x,\rho) = g_{(0)} + \,\ldots\, + \rho^{d/2} \left( g_{(d)} + h_{(d)} \log \rho \right) + \ldots,
\end{equation}
where all the terms up to the order displayed are locally determined by $g_{(0)}$ except $g_{(d)}$. $h_{(d)ij}$ is only present for even $d$. $g_{(d)}$ carries information about the state. As will become clear momentarily, in situations where AdS/CFT applies, the holographic dictionary gives $g_{(d)ij} \sim \braket{T_{ij}}$ \cite{Haro2000}.

To gain some intuition about these geometries, one can perform the coordinate transformation  $\rho =\nobreak - r^2/2$, $t=\nobreak s/r$,
with $s,r>0$, covering only the region interior to the future nullcone. The ambient metric becomes
\begin{equation} \label{AmbientMetricAdSSlicing}
    \tilde{g} = -ds^2 + s^2
    \left[ \frac{dr^2 + g_{ij}(x,r) dx^i dx^j}{r^2} \right].
\end{equation}
The Ricci-flat condition implies that the term in brackets is an asymptotically locally (Euclidean) AdS $\mbox{(ALAdS)}$ spacetime \cite{Fefferman85, Fefferman:2007rka}. The coordinate $s$ geometrises the AdS radius and thus scaling dimensions with respect to the ambient homothety $T$ coincide with engineering dimensions.  
Minkowski space can be foliated by hyperbolic slices and \eqref{AmbientMetricAdSSlicing} is the generalisation to Ricci-flat spacetimes retaining the homothety $T$. This may also lead to a connection of our work with flat-space holography, see e.g. \cite{deBoer:2003vf,Ball:2019atb,Baiguera:2022lsw}.

Weyl transformations are induced by a specific class of ambient diffeomorphisms \cite{Fefferman:2007rka} and the ambient connection $\wt{\nabla}_M$ induces a Weyl connection on the nullcone at $\rho=0$.
This is the precise sense in which Weyl transformations are realized on the ambient space, 
and CFT kinematical constraints are enforced by using appropriate diffeomorphism invariants in ambient space.
Here we restrict to non-spinning operators; spinning operators may be analysed similarly (using tractor calculus \cite{Bailey1994, vcap2003standard}).

\paragraph{Proposal.}
Our aim is to solve kinematical constraints: given a CFT with a spectrum of operators with dimensions $\Delta_i$ 
we want to determine the form of CFT correlators consistent with the kinematical constraints.
CFT correlators obey specific transformation rules under Weyl transformations and they should have the right singularity structure at short-distances, reproducing the known flat space behaviour, so we would like to obtain expressions that satisfy these two properties.

Let us first consider the case of scalar 2-point functions, generalising the embedding space result \eqref{EspaceTwopoint}. Let $X$ denote coordinates on the null cone in ambient space at $t=1$. Then $\braket{O(X_1)O(X_2)}$ is a scalar bilocal which depends on the positions of the insertions $X_1,X_2$ and has the right weight with respect to Weyl transformations. There are two main differences relative to the embedding formalism. Firstly, on a curved background $\tilde g$ it does not make sense to take an inner product between two position vectors. Secondly, $\tilde g$ in general has non-zero curvature.

Regarding the first point, on a flat ambient space the homothetic vector is given by $T = X^M\partial_M$ and thus in this case $X_{12} = -2X_1\cdot X_2$ is also equal to $X_{12} = -2T_1 \cdot T_2$. Moving away from flat space, we view $T$ as the generalisation of a position vector. To construct an inner product in this case we first parallel transport $T_1$ from $X_1$ to $X_2$ along an ambient space geodesic, giving $\hat{T}_1$ in the tangent space at $X_2$. Then we generalise $X_{12}$ by defining the dimension $-2$ scalar invariant $\widetilde{X}_{12} = -2 \, \hat{T}_1 \cdot T_2$; a short computation shows \cite{ParisiniToAppear}
\begin{equation}
\label{X12geo}
    \widetilde{X}_{12} = \ell(X_1,X_2)^2,
\end{equation}
where $\ell(X_1,X_2)$ is the geodesic length between the two points. Clearly, $\widetilde{X}_{12} = X_{12}$ in the case of flat space. Note also that although the insertions $X_1, X_2$ lie on the nullcone, the geodesics connecting them generically pass through the ALAdS slices and are thus affected by the CFT state.

Moving to the second point, non-vanishing ambient curvature entails that more scalar bilocals than $\widetilde{X}_{12}$ may be constructed. The additional ingredients are the ambient Riemann tensor $\wt{R}_{MNPQ}$, ambient covariant derivatives $\wt{\nabla}$, and $\hat{T}_1, T_2$. These ingredients are subject to the constraints $\wt{\nabla}_M T_N = \tilde{g}_{MN}$ and $T^M \wt{R}_{MNPQ}=0$. 
$\wt{R}_{MNPQ}$ encodes information about the state since for instance for even $d$, 
\be
\label{generalRiemann}
\left. (\wt{\nabla}_\rho)^{\frac{d}{2}-2} \wt{R}_{\rho ij \rho}\right|_{\rho=0,t=1}= \frac{1}{2} 
\left( \frac{d}{2}\right)! \; g_{(d)ij} + F[g_{(0)}],
\ee
where $F[g_{(0)}]$ is a local functional of the background metric $g_{(0)}$, and similarly for other components.
In holographic theories $g_{(d)ij} \sim \braket{T_{ij}}$ and when the CFT admits a large-$N$ limit $\braket{:\!T^n\!:}\sim \braket{T}^n$. In such cases products of $\wt{R}_{MNPQ}$ capture multi-stress-energy tensor contributions.

We denote ${\cal I}^{(k)}$ the general linear combination (with constant coefficients) of all curvature invariants of weight zero (w.r.t. ambient homothety) with $k$ ambient Riemann tensors.\footnote{Where sequences of covariant derivatives appear in the construction of ${\cal I}^{(k)}$, e.g. $\nabla_a \nabla_b \ldots$ acting on the same object we symmetrise their indices. This makes the counting of Riemann tensors unambiguous.} Note that depending on the example when we evaluate these invariants on the ambient background some of the terms in ${\cal I}^{(k)}$ may become linearly dependent. With all scalar bilocals in hand, we can write an expression for a general 2-point function of scalars $O$ of dimension $\Delta$ which has the right transformation properties and the correct singularity behaviour as follows,
\begin{equation}\label{eq:ProposalForm1}
	\braket{O(X_1) O(X_2)}
	=
	\frac{C_\Delta}{(\wt{X}_{12})^{\Delta}}\, \lim_{\substack{\rho\to 0\\ t\to 1}}\left[1+\sum_{k=1}^\infty\mathcal{I}^{(k)}\right] +\cdots,
\end{equation}
capturing a universal subset of all possible terms consistent with Weyl invariance, as explained below (the dots indicate that in general additional terms are present). 
We note that the 2-point function \eqref{eq:ProposalForm1} is analytic in $g_{(0)}$, as it should be. 

To get some insight in this expression let us consider some additional dynamical information. A local CFT in any state in a short-distance / high-energy limit should have an OPE expansion, which may be used to determine the correlators\footnote{Note however that {\it a priori} it appears possible to have correlators satisfying CFT kinematical constraints without necessarily having an underlying OPE, such as in the context of dS/CFT.}   (see for example \cite{Zinn-Justin:1989rgp,Fredenhagen:1986jg,Hollands:2006ag}). Therefore 2-point functions could be expressed as an expansion in terms of 1-point functions. From this perspective, the invariants ${\cal I}^{(k)}$ capture the contribution of multi-stress-energy tensors, at least for holographic CFTs in the large-$N$ limit. More generally the invariants may form a basis and \eqref{eq:ProposalForm1} may have wider applicability, as we will see in the example below.
In general the right hand side of \eqref{eq:ProposalForm1} would also contain additional terms capturing non-trivial 1-point functions of operators other than the stress-energy tensor.\footnote{We thank Slava Rychkov for discussion on this point.}  To capture these, additional invariants based on matter fields in ambient space may be required. 


In the large $\Delta$ limit we expect from usual saddle-point arguments that the 2-point function is well approximated by the geodesic length in ALAdS$_{d+1}$ connecting the two insertion points. One can show \cite{ParisiniToAppear} that geodesic lengths in ALAdS$_{d+1}$ are related to those of ambient space geodesics through
\begin{equation} \label{GeodApproxAdSAmbient}
    (\wt{X}_{12})^{-\Delta} = \left. r^{-2\Delta} e^{-\Delta L_{AdS}}\right|_{r=0},
\end{equation}
where $L_{AdS}$ is the geodesic distance between the boundary insertions on the ALAdS$_{d+1}$ slice of unit radius. Thus only the leading term in \eqref{eq:ProposalForm1} will remain in the large $\Delta$ limit. The terms in 
the sum provide the finite-$\Delta$ corrections. Using the curvature invariants of the form \eqref{CurvatureInvariantsForm2} a general result is that $\mathcal{I}^{(1)}=0$ \cite{ParisiniToAppear} and therefore the geodesic approximation is exact up to $O(\wt{\text{R}}\text{iem})^2$ assuming that no other operator with $\Delta <2d$ acquires a VEV.

As a final remark, observe that in the case of more than one ambient geodesic connecting the two insertions, new independent invariants may be associated to different geodesics, and one should sum over geodesics. Note however that under mild conditions there is a unique geodesic that connects any two points on the boundary \cite{Mazzeo86, Graham:2017fmz}. 

\paragraph{Thermal CFTs.}
We now consider the example of holographic CFTs at finite temperature living on $S^1_{\beta}\times \R^{d-1}$. We parametrize such background with coordinates $x^i=(\tau, x^a)$ with $a=2\dots d$, $\tau \sim \tau + \beta$
and denote $|x| = \sqrt{\tau^2+x^2}$. The inverse temperature $\beta$ introduces a new scale that breaks conformal invariance and thus this appears as one of the simplest settings where we can test our proposal. 
We work perturbatively in $1/\beta$ and since $\beta$ is the only scale this corresponds equivalently to a short-distance or low-temperature expansion. 

Using holography we can write down the ambient metric as \eqref{AmbientMetricAdSSlicing} where the $d+1$ dimensional metric in square brackets is given by the Euclidean AdS planar black brane, 
\begin{equation} \label{BBmetric}
  \frac{1}{z^2}\! \!\left[ 
    \frac{dz^2}{1-\frac{z^d}{z_H^d}} + \left(\! 1-\frac{z^d}{z_H^d} \right)\! d\tau^2 + \delta_{ab} dx^a dx^b
    \right]\!,
\end{equation}
where $\beta= 4 \pi z_H/ d$.
Given the ambient metric we can construct the ambient building blocks discussed above in this specific example. The use of the holographic metric here does not necessarily mean that the 2-point function constructed through \eqref{eq:ProposalForm1} using such invariants will only apply to holographic thermal CFTs; the solution to the kinematical constraints at strong coupling may provide a basis for the general solution. We will see shortly that in the present context this is indeed the case.

The ambient geodesic distance between the two insertions can be easily computed perturbatively in $1/\beta$, yielding for the first two orders when $d=4$,
\begin{align}\label{BB2ptGeodApprox4d}
		\wt{X}_{12} &=  |x|^2 \left[ 1+
		\frac{\pi ^4 |x|^2 \left(x^2-3 \tau ^2\right)}{120 \beta ^4} +  \right. \\ 
		&\quad \left. -\frac{\pi ^8 |x|^4 \left(91 \tau ^4-98 \tau ^2 x^2+19 x^4\right)}{201600 \beta ^8} + O(\beta)^{-12}
		\right],\nonumber
\end{align}
which matches the expressions for the AdS geodesic distance given in \cite{Fitzpatrick2019,RodriguezGomez2021} through \eqref{GeodApproxAdSAmbient}.

The remaining invariants can be constructed in terms of the Riemann tensor evaluated at $X_2$. We build scalar bilocals involving $k$ curvatures as follows,
\begin{equation}\label{CurvatureInvariantsForm2}
 (\hat{T}_1)^{\ell} \otimes (\wt{\nabla})^{r_1} \wt{R} \otimes \dots \otimes (\wt{\nabla})^{r_k}
 \wt{R},
\end{equation}
where $\wt{R}$ is the ambient Riemann curvature tensor at $X_2$.
Their scaling dimension is $q = 2k +r -\ell$, where $r = \sum_i^k r_i$ and $\ell$ the number of $\hat{T}_1$ vectors. One can show that $q$ is always even from geometric identities. To make a term of dimension zero which contributes to ${\cal I}^{(k)}$ we multiply by $(\wt{X}_{12})^{q/2}$.

In this example and to first order in $g_{(d)}$ (specialising \eqref{generalRiemann}), 
\begin{align}
       \wt{R}_{\rho j k\rho} &= \frac{d}{4} \left( \frac{d}{2}-1\right) g_{(d)jk} \, \rho^{\frac{d}{2}-2} t^2, \nonumber\\
	\wt{R}_{\rho jkl} &= \frac{d}{4}  \left[
	\nabla_l g_{(d)jk} - \nabla_k g_{(d)jl}
	\right] \rho^{\frac{d}{2} - 1} t^2 \label{explicitRiemann} \,, \\
        \wt{R}_{i j k l} &= \frac{d}{4} \,
        \left[ 
        g_{(0)il} g_{(d)jk} + g_{(0)jk} g_{(d)il}-(l\leftrightarrow k)
        \right] \,
        \rho^{\frac{d}{2}-1} t^2
        ,\nonumber
\end{align}
where $g_{(d)jk} \sim \braket{T_{jk}}$, the expectation value of the holographic energy momentum tensor, and $\nabla$ indicates the Levi-Civita connection associated with $g_{(0)}$. Note that $\wt{R}_{MNPQ} = O(\beta)^{-d}$ and so in general ${\cal I}^{(k)}= O(\beta)^{-kd}$. As mentioned earlier, ${\cal I}^{(1)} =0$ and the coefficient of the term at $O(\beta)^{-d}$ is fully fixed by the geodesic approximation $(\wt{X}_{12})^{-\Delta}$.
 This is in line with expectations from the thermal OPE described below and results in the literature \cite{Fitzpatrick2019, Kulaxizi:2019tkd, Karlsson:2019dbd}.

At next order $\beta^{-2d}$, there are contributions from $(\wt{X}_{12})^{-\Delta}$ and ${\cal I}^{(2)}$. To this order a complete basis is provided by terms with two Riemann tensors $\{e_0,e_1,e_2\}$, so that ${\cal I}^{(2)} = c_0 e_0 + c_1 e_1 + c_2 e_2,$ where in $d=4$,
\begin{align}
	e_0 &= \mathcal{R}^{(0)}_{AC} \, \mathcal{R}^{(0) AC}   = \frac{3}{4} \frac{|x|^8}{z_H^8} + \ldots,\nonumber\\
	e_1 &=  \mathcal{R}^{(1)}_{AC}\, \mathcal{R}^{(0)AC}   = - \frac{|x|^6}{z_H^8} (3\tau^2\!+\!7x^2)+ \ldots, \label{basisdef}\\
	e_2 &=  \mathcal{R}^{(1)}_{AC}\, \mathcal{R}^{(1)AC}  =
	4 \frac{|x|^4}{z_H^8} (3\tau^4 \!+\! 16 \tau^2 x^2\! +\! 17x^4)+ \ldots,\nonumber
\end{align}
where the ellipses denote $O(\beta)^{-12}$ corrections and
\begin{equation}\label{}
	\mathcal{R}^{(r)}_{AC} \equiv \hat{T}_1^{M_1} \dots \hat{T}_1^{M_{r}} \, \hat{T}_1^U \hat{T}_1^V \, \widetilde{\nabla}_{M_1} \dots \widetilde{\nabla}_{M_r} \widetilde{R}_{AUCV}.
\end{equation}
The polynomials on the right hand side of \eqref{basisdef} form a basis of polynomials of order 8 constructed from $x^i$ under contraction with the available boundary tensors, namely $\delta_{ij}$ and the thermal $\left<T_{ij}\right>_\beta$ appearing in \eqref{explicitRiemann}. In fact, they are linear combinations of the Gegenbauer polynomials appearing below in the discussion of thermal OPEs. Thus, in this case, as advertised, the expression of the 2-point function holds for any thermal 2-point function, even though the ambient metric was constructed using an AdS metric related to holographic thermal CFTs.

With all ambient invariants constructed to the required order in $d=4$ we have for the proposal \eqref{eq:ProposalForm1},
\begin{widetext}
\begin{align}
&\braket{O(\tau, x) O(0)}_\beta =
\frac{C_\Delta}{|x|^{2\Delta}}\left[ 1
-\frac{\pi^4 \Delta  \left(x^2-3 \tau ^2\right) |x|^2}{120 \beta^4} +\frac{\pi^8 |x|^4}{\beta^8}\left(
\left(c_0  + \frac{\Delta  (63 \Delta +170)}{30240}\right) \frac{3}{4} |x|^4\right. \right. \nonumber
\\
&\qquad\qquad\left. \left. - \left(c_1  + \frac{\Delta  (14 \Delta +39)}{25200}\right) |x|^2 (3\tau^2\!+\!7x^2) + 4 \left(c_2 + \frac{\Delta  (7 \Delta +20)}{201600}\right)  (3\tau^4 \!+\! 16 \tau^2 x^2\! +\! 17x^4) \right) + O(\beta)^{-12}\right],\label{4dprediction}
\end{align}
\end{widetext}
which is determined up to three numbers $c_0,c_1,c_2$.

As mentioned earlier we expect a short distance/high-energy expansion in any CFT in any state. For the thermal state one should expect multi-stress-energy tensor $:\!T^n(J)\!:$ contributions\footnote{In general, there are additional contributions due to other operators than the stress-energy tensor acquiring an 1-point function in the thermal state, as discussed earlier. We will not discuss these here but will return to them in \cite{ParisiniToAppear}.} of the form \cite{Katz:2014rla, Witczak-Krempa:2015pia, Iliesiu2018, Gobeil:2018fzy, Karlsson:2021duj,Fitzpatrick2019, Li:2019tpf},
\begin{equation} \label{ThermalCorrOPE}
\braket{O(\tau,x)O(0)}_\beta \supset 
    \sum_{n=0}^\infty \sum_{\substack{J=0\\J\,\text{even}}}^{2n} a^{(T)}_{n,J} \, C^{(\nu)}_{J}(\eta) \frac{|x|^{n d-2\Delta}}{\beta^{n d}}
\end{equation}
where $C^{(\nu)}_{J}$ are the Gegenbauer polynomials with order $\nu = d/2 -1$ depending on $\eta=\tau/|x|$. The constants $a^{(T)}_{n,J}$ are related to the dynamics and are not determined by symmetries. 

To the required order in $\beta$ the multi-stress-energy tensor contributions in \eqref{ThermalCorrOPE} match \eqref{4dprediction} for any $\Delta$  with the following identification,
\begin{align}
a_{0,0}^{(T)} &= C_\Delta, \qquad     a_{1,0}^{(T)} = 0, \qquad
    a_{1,2}^{(T)} = \frac{\Delta  }{120} C_{\Delta }, \label{OPEbeta4}\\
    a_{2,0}^{(T)} &= \left(\frac{3 c_0}{4}-6 c_1+52 c_2+\frac{\Delta  (7 \Delta +18)}{201600}\right) C_{\Delta },\label{OPEbeta8_1}\\
    a_{2,2}^{(T)} &= \left(c_1-15 c_2+\frac{\Delta  (7 \Delta +12)}{201600}\right) C_{\Delta },\label{OPEbeta8_2}\\
    a_{2,4}^{(T)} &= \left(c_2+\frac{\Delta  (7 \Delta +20)}{201600}\right) C_{\Delta }.\label{OPEbeta8_3}
\end{align}
The connection between the OPE thermal blocks in \eqref{ThermalCorrOPE} and ambient invariants can be understood
using factorization $\braket{:\!T^n\!:}\sim \braket{T}^n$ and the appearance of $\braket{T_{ij}}$ in \eqref{explicitRiemann}. From this point of view the connection will continue to hold to all orders $O(\beta)^{-nd}$ through an appropriate set of curvature invariants up to ${\cal I}^{(n)}$. Note that $a_{1,0}^{(T)}, a_{1,2}^{(T)}$ get contributions only from the geodesic distance in line with our earlier observation that ${\cal I}^{(1)} = 0$.

As an explicit check of the proposal \eqref{eq:ProposalForm1} in the case of thermal CFTs \eqref{4dprediction} we now compute the 2-point function using a holographic bulk computation.
We solve
\be
\Box \Phi= \Delta(\Delta-d)\Phi, \label{boxphi}
\ee
on the $d+1$ dimensional background \eqref{BBmetric} subject to Dirichlet boundary conditions and regularity in the interior. We normalise such that $C_\Delta =1$.

In the case of odd $(d+2\Delta)$ one can solve \eqref{boxphi} analytically to arbitrarily high order in $1/\beta$ in Fourier space, and for concreteness we present results for $d=4, \Delta=3/2$. We find perfect agreement with \eqref{4dprediction} which determines the following coefficients of the ambient proposal,
\begin{equation}
  \quad c_0=  -\frac{53}{1575}, 
  \quad c_1= -\frac{11}{1120},
  \quad c_2= -\frac{11}{16800}.
\end{equation}
Returning to the case of general $d$, for non-integer $\Delta$ we have also computed the order $\beta^{-d}$ contribution exactly,
\begin{equation}\label{}
	\braket{OO}_{\beta} = \frac{1}{|x|^{2\Delta}} \!\! \left[
	1 + \lambda_1\!
	\left(x^2\!-\!(d-1) \tau ^2\right)\!
	\frac{|x|^{d-2}}{\beta^d}
	\right] + O\!\!\left(\beta\right)^{-2d}\!,
\end{equation}
where $\lambda_1 = \left(\frac{4\pi}{d}\right)^d\frac{\sqrt{\pi } (-1)^{d+1} \Delta   \, \Gamma \left(-\frac{d}{2}-\frac{1}{2}\right) \sin (\pi  (d-\Delta ))}{2^{d+2} \Gamma \left(1-\frac{d}{2}\right) \tan \left(\frac{\pi  d}{2}\right) \sin (\pi  \Delta ) }.$

\paragraph{Squashed spheres.} We now present an example where the result is not known and not easy to obtain by any other means: the 2-point function of scalar operators for CFTs on squashed spheres. See \cite{Zoubos:2002cw,Zoubos:2004qm} for a holographic computation on such backgrounds.
 We fix $d=3$ for concreteness; the background takes the form 
\begin{equation}\label{}
	ds^2 = d\theta^2 +\sin^2\theta d\phi^2 + \frac{1}{1+\alpha}  \left( d\psi + \cos\theta d\phi\right)^2,
\end{equation}
where $\alpha$ parametrises the squashing and $\alpha=0$ corresponds to a round $S^3$. Here $0\leq \psi < 4 \pi$, $0<\theta<\pi$ and $0<\phi<2\pi$. The CFT state that we intend to study on this background is characterised by the stress tensor VEV $\braket{T_{ij}} = \frac{3}{16 \pi} g_{(3)ij}$, with
\begin{equation}
	\begin{split}
			&g_{(3)ij}(x) dx^i dx^j = \frac{\alpha }{3 (\alpha +1)^{3/2}}
	\left[
	d\theta^2 -\frac{2}{1+\alpha}d\psi^2  \right.
 \\& \left.-\frac{4 \cos\theta}{1+\alpha} d\psi d\phi - \frac{(\alpha +3) \cos 2 \theta -\alpha +1}{2 (\alpha +1)} d\phi^2
	\right].
	\end{split}
\end{equation}
We work perturbatively in small $\alpha$. According to our prescription, the ALAdS$_4$ slices of the ambient space corresponding to this setup are AdS-Taub-NUT$_4$ spaces \cite{stephani_kramer_maccallum_hoenselaers_herlt_2003},
\begin{equation}\label{}
	\frac{dr^2}{V(r)}
	+ (r^2-n^2) (d\theta^2 +\sin^2\theta d\phi^2) + 4 n^2V(r) \left(
	d\psi + \cos \theta d\phi	\right)^2
\end{equation}
with $n= (2\sqrt{\alpha +1})^{-1} $, $m = \alpha (1+\alpha)^{-3/2}/2$ and
\begin{equation}\label{}
	V(r) = \frac{r^2 + n^2 -2 m r+\left(r^4 -6 n^2 r^2 -3 n^4\right)}{r^2 - n^2}.
\end{equation}
We consider insertions at the generic points $x_1 = (\theta_1, 0,0)$ and $x_2=(\theta, \psi, \phi)$ using translation invariance along $\psi$ and $\phi$. We further fix $\theta_1=0$ for ease of exposition. 

Defining $\chi=(\phi -\psi) /2$, the parallel transported vector from $X_1$ to $X_2$ reads
\begin{align}
		&\hat{T}_1 = \frac{\sin \frac{\theta }{2} \cos \chi+1}{2 r} \p_s + \frac{\sin \frac{\theta }{2} \cos \chi-1}{64 r} \p_r \\ 
  \!\!&+ 2 \cos \frac{\theta }{2} \cos \chi \p_\theta 
		+ \csc \frac{\theta }{2} \sin \chi\p_\psi -\csc \frac{\theta }{2} \sin \chi\p_\phi + O(\alpha). \nonumber
\end{align}
The non-vanishing components of the ambient Riemann are of the form \eqref{explicitRiemann},  with the addition of an overleading $( \nabla_l R_{(0)jk} - \nabla_k R_{(0)jl}) t^2$ to 	$\wt{R}_{\rho jkl}$.

Following the same arguments as in the thermal case, $\mathcal{I}^{(1)}=0$ and the leading curvature invariants are thus of order $O(\alpha)^2$. A possible basis for the ambient invariants accounting for the three $:\! T^2 \! :$ blocks including no derivatives is given by (up to order $O(\alpha)^3$),
	\begin{align}
	(\nabla \wt{\text{R}}\text{iem})^2  &= \frac{168 \alpha^2}{t^6}, \\
	\wt{\mathcal{R}}^{(1)}_{AC}\, \wt{\mathcal{R}}^{(1)AC} &=
	18 \alpha ^2 \cos ^2\!\frac{\theta }{2} \left[\cos \theta +3-2 \sin ^2\!\frac{\theta }{2} \cos 2 \chi \right]^2, \nonumber\\
	\wt{\mathcal{R}}^{(0)}_{AC}\, \wt{\mathcal{R}}^{(2)AC} &=
	-\frac{3}{2}  \alpha ^2 \left(\cos \theta +3 -2 \sin ^2\!\frac{\theta }{2} \cos 2 \chi \right)^2 \nonumber \\
 &  \qquad \qquad \quad \left( 5 \cos \theta +3 
+ 2 \sin ^2\!\frac{\theta }{2} \cos 2 \chi \right),
 \nonumber
\end{align}
since from the form of the ambient Riemann it follows they do not contain derivatives acting on the stress tensor. A linear combination of these three invariants with general coefficients will appear in $\mathcal{I}^{(2)}$. Similar invariants can be constructed for  double-stress tensor blocks with an arbitrary number of derivatives as well as for multi-stress-energy tensor blocks. Using these ingredients, we can assemble the correlator according to \eqref{eq:ProposalForm1} (more details will be provided in \cite{ParisiniToAppear}).

\paragraph{Higher-point functions. }
Following similar arguments, higher point functions also admit an ambient space representation,
\begin{equation}
    \braket{O_1(X_1) \ldots O_n(X_n)} =  \left(\prod_{\text{pairs}} \widetilde{X}_{ij}^{\alpha_{ij}}\right) \left[f\left(u\right)+\ldots\right],\label{higherpointansatz}
\end{equation}
where ellipses denote weight-zero ambient curvature invariants akin to those built for 2-points. Here $\widetilde{X}_{ij}$ are ambient geodesic distances \eqref{X12geo} connecting all pairs of insertions. $f$ is the function of cross ratios that appears for $n$-point functions of the same CFT in vacuum on flat space, where now the cross-ratios are formed from the geodesic pairs, $u_{[pqrs]} = (\widetilde{X}_{pr}\widetilde{X}_{qs})/(\widetilde{X}_{pq}\widetilde{X}_{rs})$ and where $\Delta_i = -\sum_{j=1}^n\alpha_{ij}$. 
As discussed earlier, in many cases ambient invariants involving a single Riemann tensor vanish, and thus \eqref{higherpointansatz} may be computed using a geodesic approximation to first subleading order in the deviation from flat space vacuum CFT. 

\paragraph{Conclusions.}
We have presented a prescription to solve the kinematical constraints of scalar $n$-point functions for CFTs in general backgrounds and states.
The construction is based on a generalisation of the embedding space formalism and utilises geometric invariants of the ambient space.
Our proposal captures stress-energy tensors; to capture the contributions of other operators the ambient space should be generalised to include matter fields.
We tested the construction in the case of holographic 2-point functions for CFTs in a thermal state finding exact agreement, and along the way confirmed expectations from thermal OPEs. We made predictions for 2-point functions on squashed spheres and for general scalar $n$-point functions. 
Natural generalisations of this proposal include higher-spin operators leveraging tractor calculus.

\begin{acknowledgments}
{\em Acknowledgments.} We thank Slava Rychkov for comments. The work of EP is supported by the Royal Society Research Grants RGF/EA/181054 and RF/ERE/210267. KS and BW are supported in part by the Science and Technology Facilities Council (Consolidated Grant “Exploring the Limits of the Standard Model and Beyond”). BW is supported by a Royal Society University Research Fellowship.
\end{acknowledgments}

\bibliography{BIB} 

\end{document}